\documentclass[aps,pra,showpacs,preprint,superscriptaddress]{revtex4-1}

\usepackage{amssymb}
\usepackage{graphicx}
\usepackage{bm}
\usepackage{amsfonts}
\usepackage{color}
\usepackage[usenames]{xcolor}

\begin{document}

\title{Localization of the valence electron of endohedrally confined hydrogen, lithium and sodium in fullerene cages}

\author{Eloisa Cuestas}
\email{mecuestas@famaf.unc.edu.ar}
\affiliation{Facultad de Matem\'atica, Astronom\'{\i}a y F\'{\i}sica,
Universidad Nacional de C\'ordoba and IFEG-CONICET, Ciudad Universitaria,
X5016LAE C\'ordoba, Argentina}

\author{Pablo Serra}
\email{serra@famaf.unc.edu.ar}
\affiliation{Facultad de Matem\'atica, Astronom\'{\i}a y F\'{\i}sica,
Universidad Nacional de C\'ordoba and IFEG-CONICET, Ciudad Universitaria,
X5016LAE C\'ordoba, Argentina}

\begin{abstract}

The localization of the valence electron of $H$, $Li$ and $Na$ atoms enclosed by three different fullerene molecules is studied. The structure of the fullerene molecules is used to calculate the equilibrium position of the endohedrally atom as the minimum of the classical $(N+1)$-body Lennard-Jones  potential. Once the position of the guest atom is determined, the fullerene cavity is modeled by a short range attractive shell according to molecule symmetry, and the enclosed atom is modeled by an effective one-electron potential. In order to examine whether the endohedral compound is formed by a neutral atom inside a neutral fullerene molecule $X@C_{N}$ or if the valence electron of the encapsulated atom localizes in the fullerene giving rise to a state with the form $X^{+}@C_{N}^{-}$, we analyze the electronic density, the projections onto free atomic states, and the weights of partial angular waves.
\end{abstract}

\keywords{endohedral compounds; fullerene; valence electron.}

\date{\today}

\maketitle

\section{Introduction}

Fullerenes are stable carbon molecules with spherical, ellipsoidal or cylindrical hollow shape. The existence of these carbon structures was theoretically predicted by different authors between 1965 and 1975 \cite{schultz1965,osawa1970,bochvar1973}, but it was not until 1985 that Kroto \textit{et al.} observed $C_{60}$ \cite{kroto1985,kroto1987,kratschmer1990,halford2006}.

Fullerene molecules, together with graphene and carbon nanotubes, are of considerable interest to researchers from many scientific areas and have multiple possible applications, specially on nanotechnology. Their unique mechanical and electronic properties, such as low weight, high strength, flexibility and thermal stability, make this carbon structures quite suitable  for the creation of nano-devices \cite{cox2007,cox2007_2}.

One of the most important features of fullerene molecules is that they can harbor atoms in their interior constituting endohedral compounds, denoted by $X@C_{N}$ \cite{laasonen1992,bethune1993,murry1994,becker1996,shionhara2000,forro2001,averbukh2006,xu2009}, whose main advantage is the isolation of the enclosed atom from its environment. Endohedral atoms can be produced in laboratories  \cite{yannoni1992,kato1993,saunders1993,saunders1996,murphy1996,becker2000,komatsu2005,dunka2013}, and many technological applications, such as drugs-delivery agents, molecular containers,  hydrogen or lithium storage, are now in development \cite{mitnik2008}.

Previous research mostly aimed to investigate the shell structure and spectral properties of $X@C_{60}$, in which the enclosed atom is located right at the geometrical center of $C_{60}$, and the fullerene molecule is modeled by a short-range spherical shell with an attractive potential \cite{mitnik2008,xu1996,connerade1999,cdm99,decleva1999,quiao2002,amal2007,ndengue2009,lin2012}. Nevertheless, for larger fullerene molecules the equilibrium position of the confined atom is no longer the geometrical center of the molecule \cite{shi2000,kang2006}.

The purpose of the present work is to calculate and describe the physically expected differences between the electronic states of the atom confined by the fullerene cage and their free state. We focus on the systems $X@C_{N}$, where $X=H$, $Li$, $Na$, and $N=80,\, 180$. The chosen $C_{80}$ isomers were $C_{80}-D2-2$ and $C_{80}-D5d-1$ because they are the only two isomers that have been prepared and characterized as pristine structures \cite{rios2012}. 

Each of these fullerene molecules was modeled by a short range shell according to molecule symmetry. $C_{180}$ was modeled by a spherical shell, while both $C_{80}$ isomers were described by an ellipsoidal shell with proper parameters. 

The fundamental question underlying our work could be stated in the following way: is the endohedral compound formed by a neutral atom inside a neutral fullerene ($X@C_{N}$), or does the valence electron of the encapsulated atom localize into the fullerene shell, giving rise to a zwitterion-like state ($X^{+}@C_{N}^{-}$)?

In order to understand the transference of the valence electron to the fullerene cage $X@C_{N} \, \rightarrow \, X^{+}@C_{N}^{-} \,$, we analyze the electronic density, the projections onto free atomic states, and the weights of partial angular waves as a function of the attraction that the fullerene exerts upon the valence electron. 

It is important to mention that the short range spherical (or ellipsoidal) potential for the fullerene cage is indeed a very simple approach; even so, this model explains and provides a good description of the system and its properties. There are plenty of studies of the electronic structure as well as photoionization of $X@C_{60}$ where the fullerene molecule is modeled as a spherical short range shell \cite{dolmatov2009,korol2010,dolmatov2012,gorczyca2013,dolmatov2014}, which show an excellent agreement with another theoretical studies as well as with the experimental photoionization cross section or electron elastic scattering cross section \cite{cioslowski1991,nascimento2011,baltenkov1999,madjet2010,lin2013}.

This paper is organized as follows. In the second section, we present a complete description of the theoretical method; the third section is focused on the computation, results and discussion. Finally, our conclusions are given in the fourth section. 

\section{Model and theoretical methods}
\label{model}

In a first step we used the structure of the fullerene molecules to obtain the position of the confined atom as the minimum of the classical $(N+1)$-body potential as explained in Sec.  \ref{obtained X-position}. Once the position of the atom is determined, the electronic structure of the enclosed atom is calculated modelling the fullerene as a spherical or ellipsoidal square well according to the symmetry of the fullerene molecule.

\subsection{Determination of the position of the confined atom}
\label{obtained X-position}

We model the interaction between the $X$ atom and the carbon fullerene using the non-bonded classical Lennard-Jones potential with each carbon atom of the $C_{N}$.

\begin{equation}
\label{eljp}
V_{N}^{(X)}(\vec{x})\,=\,\sum_{i=1}^N\,v_{(X)}(|\vec{x}-\vec{x}_i|) \;\;\;;\;\;\;
v_{(X)}(r)=4 \epsilon_{(X)}\left(\left(\frac{\sigma_{(X)}}{r}\right)^{12}-\left(
\frac{\sigma_{(X)}}{r}\right)^6 \right)\,,
\end{equation}

\noindent where $\vec{x}$ is the position of the confined atom, $\vec{x}_i$ the position of the $i$-th carbon from Ref. \cite{tomanek14}, and the pair $\epsilon_{(X)}$, $\sigma_{(X)}$ are the Lennard-Jones parameters adjusted for the $X-C$ interaction, displayed in Table \ref{tljp} ($au$ are used throughout this paper). The equilibrium position $\vec{x}_{0}$ of the $X$ confined atom is calculated minimizing this  potential,

\begin{equation}
\label{eljm}
V_{N}^{(X)}(\vec{x}_{0})\,=\,\min_{\{\vec{x}\}} V_{N}^{(X)}(\vec{x})\,.
\end{equation}

\begin{table}[ht]
\begin{tabular}{|c|c|c|c|}\hline\hline
\hspace{.5cm}  $X$ \hspace{.5cm} &  \hspace{.5cm} $\epsilon_{(X)}$ \hspace{.5cm}  &
\hspace{.5cm} $\sigma_{(X)}$ \hspace{.5cm} \mbox{} \\ \hline
$H^{(a)}$  &   $ 8.7488 \; 10^{-5}$ & $5.820$\\ \hline
$Li^{(b)}$  &  $ 1.5921 \; 10^{-4}$  &  $4.668$\\ \hline
$Na^{(c)}$  &  $ 6.574 \; 10^{-5}$  &  $5.990$
 \\ \hline\hline
\end{tabular}
\caption{\label{tljp}  Lennard-Jones parameters for $X-C$ interactions. (a) from Ref. \cite{xsblt08}, (b) from Ref. \cite{chan2011} and (c) from Ref. \cite{chan2012}.}
\end{table}

\begin{figure}
\begin{center}
\includegraphics[width=6.cm]{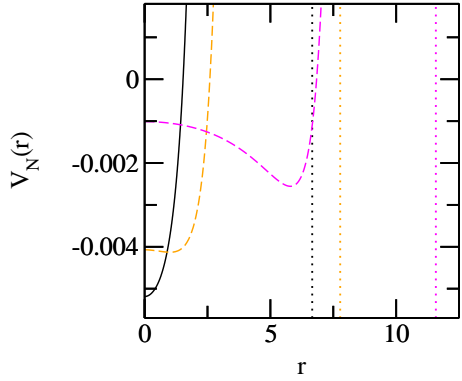}
\end{center}
\caption{  \label{fljp} (color online) LJ potential Eq.(\ref{eljp}) for the hydrogen
atom inside a $C_N$ for $N=60$ (black line), $N=80$ (orange line) and $N=180$ (magenta line). The average sizes of the respective fullerenes are shown in
dashed lines with the same set of colors.   
}
\end{figure}

In Fig.  \ref{fljp} we present, for the hydrogen case, the LJ potential $V_{N}^{(H)}$ calculated along a line between the center of the $C_{N}$ molecule and one of its carbon atoms. The calculations were made with $N=60,80, 180, 320$, and the locations of the carbon atoms were fixed according to the structure of the fullerene molecules. The figure shows that only for $C_{60}$ we obtained $\vec{x}_{0}=0$ as the position of the  equilibrium minimum. For larger fullerenes, the confined atom is no longer located at the geometrical center of the fullerene.

\subsection{ Electronic States}

Most of the studies on $X@C_{N}$ have been done with $N=60$ and the endohedral atom located at the center of the fullerene modeled by a spherical potential \cite{mitnik2008,xu1996,connerade1999,cdm99,decleva1999,quiao2002,amal2007,ndengue2009,lin2012}. However, as it is shown in Fig.  \ref{fljp}, for $N>60$  the equilibrium position of the confined atom is not the fullerene center. In these cases the system is not spherically symmetric, therefore the Schr\"odinger equation  is not separable into radial and angular variables.

The system $X@C_{N}$ was modeled using the following one-electron Hamiltonian 

\begin{equation}
\label{eoeh}
H_X\,=\,-\frac{1}{2} \nabla^2 + V_X(\vec{x})+V_F (\vec{x})  \;\; ,
\end{equation}

\noindent where the potential $V_F$ models the fullerene-electron interaction, and the potential $V_X$ represents the interaction of the valence electron with  the multi-electron core. For $X=H,Li,Na$ we used a one-electron potential model for their valence electron, which includes the average effects of the inner electrons \cite{schweizer1999,sh00,sahoo2002,sahoo2005,sh06,kar2005,kar2012,kar2013}, 

\begin{equation}
\label{evep}
V_{X}(r)\,=\, 
-\frac{1}{r} \left\lbrace 
\left( Z - N_c \right) +  
N_c \left(   e ^{-2 \alpha  r}  + \beta \; r\; e ^{-2 \gamma r} \right)  \right\rbrace  
\end{equation}

\noindent  where $r$ is the distance between the outer electron and the nucleus. The fitting parameters (see Table \ref{tamp}) were taken from Ref. \cite{sh06}, in which the authors chose them in order to reproduce the best values of the energy levels of the free atom compared with experimental results. In the $Li$ case  the potential parameter $\alpha= \beta=\gamma$ describes the effective charge of the $1s^{2}$ core. Notice that for $X=H$, $N_c=0$ and thus $V_H$ reduces to the Coulombic electron-proton interaction.

\begin{table}[ht]
\begin{tabular}{|c|c|c|c|c|}\hline\hline
\hspace{.5cm}  $X$ \hspace{.5cm} &  \hspace{.5cm} $H$ \hspace{.5cm}  &
\hspace{.5cm} $Li$  \hspace{.5cm} &  \hspace{.5cm} $Na$ \hspace{.5cm}  \\ \hline
$Z$  &   $ 1$ & $3$ & $11$\\ \hline
$N_c$  &   $ 0$ & $2$ & $10$\\ \hline
$\alpha$  &   $ 0$ & $1.6559$ & $1.8321$\\ \hline
$\beta$  &   $ 0$ & $1.6559$ & $1.0591$\\ \hline
$\gamma$  &   $ 0$ & $1.6559$ & $1.3162$
 \\ \hline\hline
\end{tabular}
\caption{\label{tamp} $H$, $Li$ and $Na$ parameters for the valence electron potential Eq. \ref{evep} from Ref. \cite{sh06}.\\}
\end{table}

The fullerene-valence electron interaction potential $V_F $ is modeled by a short-range shell. The spherical shell takes the form

\begin{equation}
\label{efpotsph}
V^{spherical}_F(\vec{x})\,=\,\left\{ \begin{array}{rl}
-U_0 & \mbox{if} \;\; R - \Delta <r < R + \Delta \\
0 \;\; &\mbox{otherwise }
 \end{array}  \right. \,,
\end{equation}

\noindent where $R$ is the average radial position of the carbon atoms in the fullerene, $2 \Delta$ is the thickness of the cage shell, and $U_0$ its effective depth.

The ellipsoidal $C_{80}$ isomers were modeled as an ellipsoidal square well,

\begin{equation}
\label{efpotellip}
V^{ellipsoidal}_F(\vec{x})\,=\,\left\{ \begin{array}{rl}
-U_0 & \mbox{if} \;\; R(\theta) - \Delta <r <R (\theta) + \Delta  \\
0 \; \; &\mbox{otherwise }
 \end{array}  \right. \,,
\end{equation}

\noindent where $R(\theta) = \left\lbrace \frac{ \cos^{2} (\theta) }{a^{2}}+\frac{\sin^{2} (\theta)}{b^{2}} \right\rbrace ^ {-\frac{1}{2}}$ for the polar angle $ 0 \leq \theta \leq \pi$ (spherical coordinates), $a$ and $b$ are the ellipsoid semi-axes, and $\Delta$ and $U_0$ were defined above. The values of these parameters are shown in Table \ref{tsph-ellipp}. In all cases $2\Delta = 1.89$ from Ref. \cite{xu1996}.

\begin{table}[ht]
\begin{tabular}{|c|c|c|c|c|}\hline\hline
\hspace{.5cm}  $C_{N}$ \hspace{.5cm} &  \hspace{.5cm} $C_{80}-D2-2$ \hspace{.5cm}  &
\hspace{.5cm} $C_{80}-D5d-1$  \hspace{.5cm} &  \hspace{.5cm} $C_{180}-0$ \hspace{.5cm}  \\ \hline
$R$  &   $ 7.540$ & $7.554$ & $11.305$\\ \hline
$a$  &   $ 8.84$ & $9.19$ & $1$\\ \hline
$b$  &   $ 6.94$ & $6.74$ & $1$
 \\ \hline\hline
\end{tabular}
\caption{\label{tsph-ellipp} Spherical and ellipsoidal parameters for $C_{N}$.}
\end{table}

Our workhorse for the calculation of the electronic states is the Ritz-variational method with real square-integrable, compact support, and boundary conditions adaptable basis sets. Since the Hamiltonian is invariant in respect to rotation about z-axis, the wavefunction can be written as

\begin{equation}
\label{epsi}
\Psi(\vec{x}) \,=\, \frac{u (r, \theta)}{r} \; \Phi_{L_z} \left( \phi \right) \,,
\end{equation}

\noindent where $\Phi_{L_z}$ is a normalized eigenfunction of $L_z$, the z-component of angular momentum. We used B-Splines to expand $u (r, \theta)$,

\begin{equation}
\label{eu_BS}
u (r, \theta) \, =\, \sum_{i,j} c_{i,j} \; B_{i,k}(r) \; B_{j,\tilde{k}}(\theta)   \,,
\end{equation}

\noindent $B_{i,k}(r)$, $B_{j,\tilde{k}}(\theta)$ are radial and angular B-Splines polynomial of order $k$ and $\tilde{k}$, defined in the intervals $\left[ 0, r_{max} \right] $ and $\left[ 0, \pi \right] $ respectively. Once we defined the basis-set, the Schr\"odinger equation can be expressed in matrix form $H\,\vec{c} \,=\, E \, S\, \vec{c} $, where $E$ is the energy, $\vec{c}\,$ is the coefficient vector of the electronic state, and $S$ the basis overlap matrix. This method provides the electronic structure (orbitals and energy levels) by diagonalizing the Hamiltonian matrix.

The numerical results  are obtained by defining a cutoff radius $r_{max}$, and then the interval $[0,r_{max}]$ is divided into $I$ equal subintervals. B-spline polynomials \cite{deboor,brandefelt2001} are piecewise polynomials defined by a sequence of  knots $t_1=0\leq t_2\leq\cdots \leq t_{2 k+I-1}=r_{max}$ and the recurrence relations

\begin{equation}\label{bs1}
B_{i,1}(r)\,=\,\left\{ \begin{array}{ll} 1 & \mbox{if}\,t_i\leq r <
t_{i+1}   \\
0 &\mbox{otherwise,}  \end{array}  \right. \,.
\end{equation}

\begin{equation}\label{bsrr}
B_{i,k}(r)\,=\,\frac{r-t_i}{t_{i+k-1}-t_i}\,B_{i,k-1}(r)\,+\,
\frac{t_{i+k}-r}{t_{i+k}-t_{i+1}}\,B_{i,k-1}(r)\; (k>1)\,.\\ 
\end{equation}

In our calculations we have used seventh-order B-Splines in both variables. The sequences of knots  were chosen in order to suit the boundary conditions of the problem. In the angular variable the knots sequences have $\tilde{k}$ multiple knots in $\theta\,=\,0,\pi$. In the radial variable the knots sequences have $k$ multiple knots in $r\,=\,0,r_{max}\;$; $k-2$ multiple knots in the atom position (cusp-condition) and only in the case of the spherical model, $k-3$ multiple knots in the internal an external radius of the spherical shell. 

It is worth mentioning that in the angular variable the sequences of knots have a Gauss-Legendre-quadrature-points distribution, whose advantages have been previously discussed in the literature \cite{shi2000,kang2006}.

\section{Results and discussion}


In order to obtain the position of the confined atom we minimized the interaction potential between the $X$ atom and the carbon atoms of the fullerene molecule Eq. (\ref{eljp}), where the locations of the carbon atoms were fixed according to the structure of each fullerene from Ref. \cite{tomanek14}.

Because of the well defined spherical shape of $C_{180}-0$, the position of the endohedral atom is the Euclidean norm of the absolute minimum vector $\vec{x}_{0}$ obtained by Eq. (\ref{eljm}).  

$C_{80}-D2-2$ and  $C_{80}-D5d-1$ exhibit ellipsoidal shape, slightly broken by the discrete nature of the fullerene. Thus, instead of the expected two degenerate global minima along the major axis of the molecule for the ellipsoidal symmetry, the $(N+1)$-body potential presents a global and a second quasi-degenerated minima along the major axis. The equilibrium positions $\vert \vec{x}_{0} \vert$ for each case are presented in Table \ref{tD}. 

\begin{table}[ht]
\begin{tabular}{|c|c|c|c|c|}\hline\hline
\hspace{.5cm}  $X$ \hspace{.5cm} &  \hspace{.5cm} $C_{80}-D2-2$ \hspace{.5cm}  &
\hspace{.5cm} $C_{80}-D5d-1$  \hspace{.5cm} &  \hspace{.5cm} $C_{180}-0$ \hspace{.5cm}  \\ \hline
$H$  &   $ 1.214$ & $1.386$ & $5.310$\\ \hline
$Li$  &   $ 3.536$ & $3.837$ & $6.622$\\ \hline
$Na$  &   $ 0.321$ & $0.381$ & $5.127$
 \\ \hline\hline
\end{tabular}
\caption{\label{tD} Equilibrium position $\vert \vec{x}_{0} \vert$ of $X@C_{N}$.}
\end{table}


Once the equilibrium position of the enclosed atom was determined, we focused on the valence electron states. In order to study how the spectral properties of $H$, $Li$ and $Na$ atoms are modified by the carbon cage, our calculations included values of the depth of the attractive shell $U_0$ in the $[0,1.5]$ range.

\begin{figure}
\begin{center}
\includegraphics[width=\textwidth]{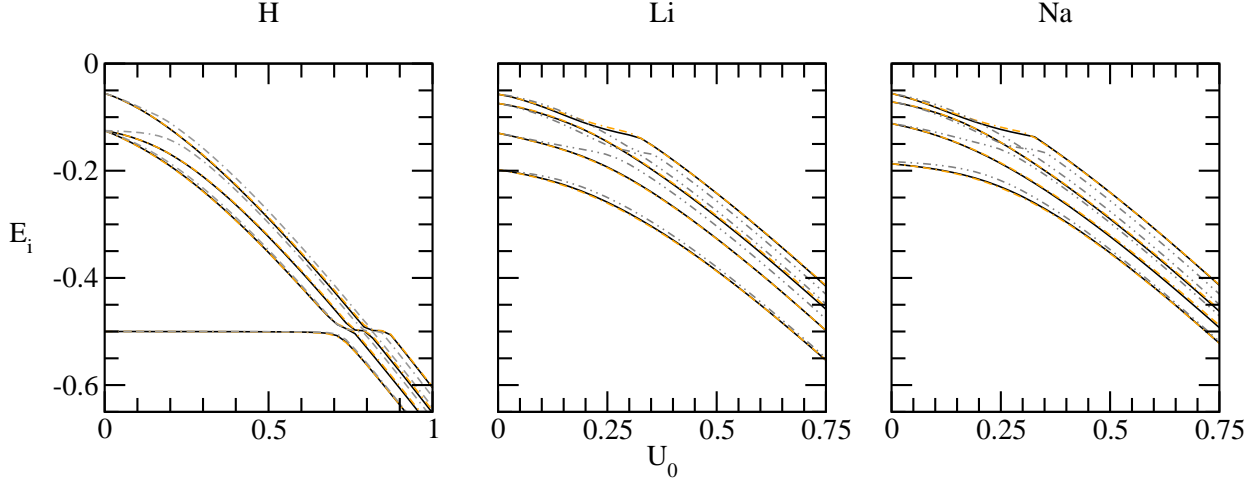}
\end{center}
\caption{  \label{av} (color online) First four energy levels of the encapsulated atoms at $C_{80}-D2-2$ (black full line), $C_{80}-D5d-1$ (orange dash-dotted line), and $C_{180}-0$ (gray dash-dotted line). 
}
\end{figure}

The first four energy levels of the encapsulated atoms for each fullerene are depicted in Fig. \ref {av} as a function of $U_0$, $C_{80}-D2-2$ in full black line, $C_{80}-D5d-1$ in dash-dotted orange line, and $C_{180}-0$ in gray dash-dotted line.

Even when the main spectrum difference is between the $C_{80}$ isomers and $C_{180}-0$, this difference is small because the distance between the core atom $X$ and the fullerene shell is  determinated by $\sigma_{(X)}$, and it is almost independent of the size of $C_N$  (see Tables \ref{tljp}, \ref{tsph-ellipp}, and \ref{tD}).

Notice that when varying the depth of the shell, even for values of $U_0 > 0.6$, the ground state of $H$ remains stable at the ionization energy of the free atom ground state $E_{0}=-0.5\, au$, a behaviour that has also been observed for the $C_{60}$ case in Ref. \cite{amal2007}. In other words, given that the ground electronic state of hydrogen is highly localized around the nucleus, it is not influenced by the outer confining well unless the fullerene becomes sufficiently attractive. The other $H$ states, as well as the $Li$ and $Na$ states, are influenced by the attractive fullerene shell even for small $U_{0}$ values because all of these states are delocalized states. 

The Ritz-variational method provides not only the eigenvalues of the Hamiltonian matrix, but also the eigenfunctions. The probability density of finding the valence electron around a position given by $r,\theta$ is proportional to $r^{2} \rho (r,\theta) \, dr \, d\theta = u^{2} (r,\theta) \, dr \, d\theta$, where $u(r,\theta)$ is defined in Eq. (\ref{epsi}). 

\begin{figure}
\begin{center}
\includegraphics[width=\textwidth]{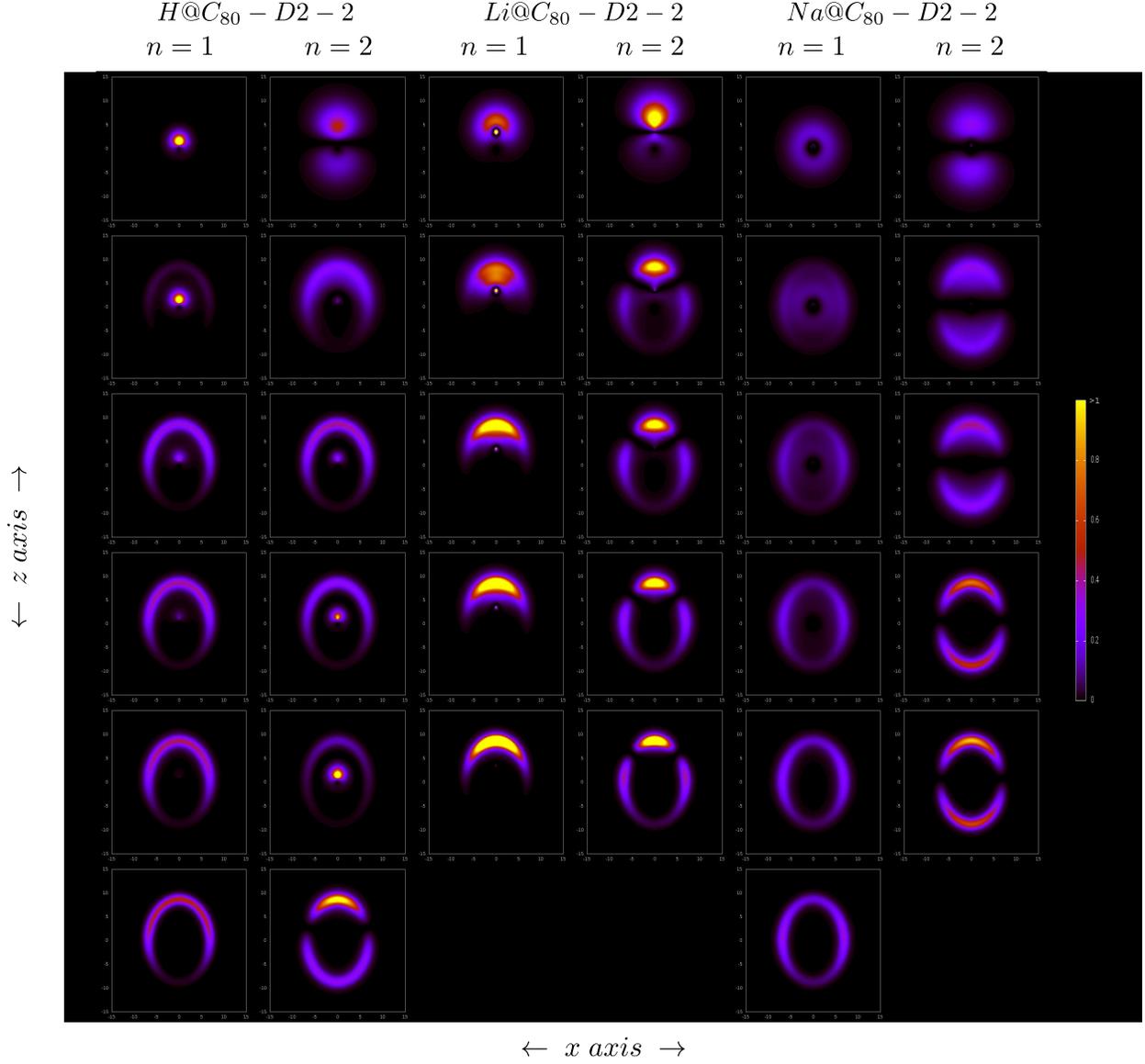}
\end{center}
\caption{  \label{fu2_D2} (color online) Valence electron probability density on $xz$ plane of the ground ($n=1$) and first excited wavefunctions ($n=2$) of $X@C_{80}-D2-2$ for increasing $U_{0}$ values from top to bottom. For $H@C_{80}-D2-2$, $n=1$ the values are $U_0 = 0, 0.682, 0.736, 0.764, 0.818, 1.364$, for $n=2$ the values are $U_0 = 0, 0.245, 0.682 , 0.709, 0.736 , 0.764 $. For $Li@C_{80}-D2-2$, $n=1$ the values are $U_0 = 0, 0.191, 0.409 , 0.491 , 0.955 $, for $n=2$ the values are $U_0 = 0, 0.273 , 0.382 , 0.545 , 0.955 $. For $Na@C_{80}-D2-2$, $n=1$ the values are $U_0 = 0, 0.191 , 0.273 , 0.355 , 0.545 , 0.818$, for $n=2$ the values are $U_0 =  0, 0.191 , 0.273 , 0.682 , 0.955$. 
}
\end{figure}

The probability of finding the valence electron on the $xz$ plane for the ground ($n=1$) and first excited wavefunctions ($n=2$) of $X@C_{80}-D2-2$ are depicted in Fig.  \ref{fu2_D2}, where the enclosed atom is located in the positive $z$-axis at the positions given in Table \ref{tD}. 

This figure shows several interesting features. First of all, the symmetry of the localization in the fullerene shell depends both on the distance between the enclosed atom and the geometrical center of the molecule (origin of coordinates) and on the symmetry of the atomic free ($U_{0}=0$) states ($s$-wave for ground state and $p$-wave for first excited state). 

It is physically expected that an electron in an $s$-wave with no spatial-direction preference will be localized in the nearest part of the fullerene shell. In the case of an electron in a $p$-wave with spatial preference for the $z$ positive and negative axis, denoted by $z+$ and $z-$, this spatial preference and the tendency to localizing in the nearest part of the fullerene attractive shell will compete. The decisive variable in this competition is the distance between the enclosed atom and the geometrical center of the fullerene.

In the case of $Na$, whose nucleus is located closer to the geometrical center of the fullerene, the valence electron in the ground state will localize in the whole fullerene shell when the fullerene gets increasingly attractive. An electron in the first excited state of $Na$ will localize mostly on the upper and lower ``hemispheres" of the fullerene shell, in a $p$-wave symmetric shape.

In the case of $Li$, whose nucleus locates the farthest from the geometrical center of the fullerene, the valence electron in the ground state will localize on the nearest part of the fullerene shell, which is the upper ``hemisphere". An electron in the first excited state of $Li$ will also localize on the lower ``hemisphere" due to the $p$-wave symmetry of the initial state ($U_{0}=0$).

As mentioned above, the ground $H$ state remains unchanged until the fullerene shell becomes sufficiently attractive ($U_{0} \sim 0.73$), and for $U_{0}$ values above this critical value, the electron becomes localized on the fullerene shell. 

The first excited $H$ state presents a quite interesting phenomenon which coincides with $U_{0}$ values for the anticrossing energy levels, $U_{0} \sim  [0.6,0.76]$ in Fig. \ref{av}. By increasing the $U_{0}$ values in the range $U_{0} \sim  [0,0.6]$ the electron becomes localized in the fullerene, and for the anticrossing values $U_{0} \sim  [0.6,0.76]$ the electron is confined into the atom and relocalized in the fullerene with a new symmetry. A similar effect was originally described and termed \textit{mirror collapse} by Connerade \textit{et al.} for the $C_{60}$ case in Ref. \cite{connerade1999}.

\begin{figure}
\begin{center}
\includegraphics[width=\textwidth]{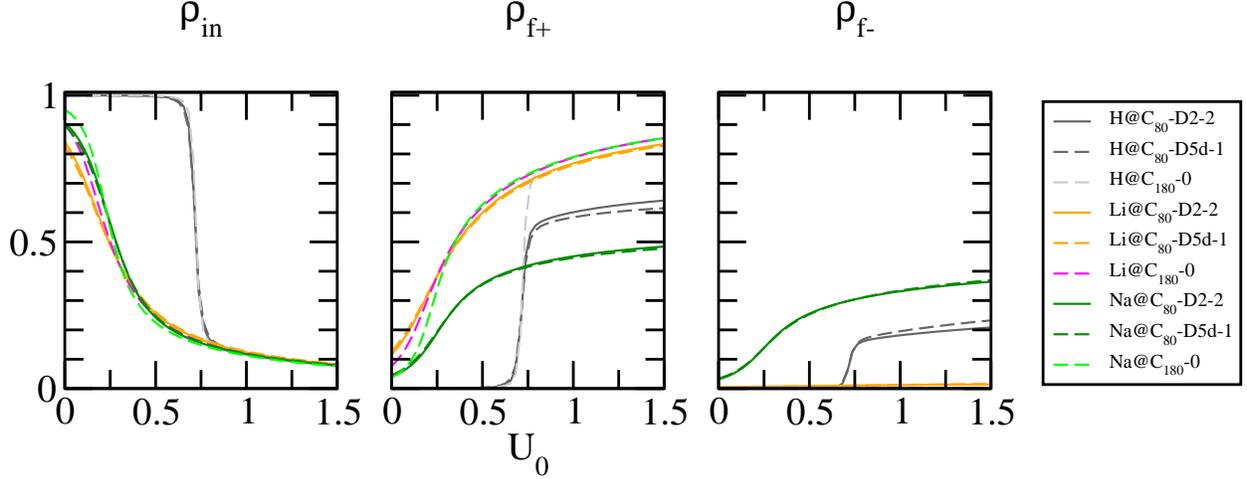}
\end{center}
\caption{  \label{frhon1} (color online) Probability of finding the valence electron in the ground state, inside the fullerene cage $\rho_{in}$ (first column from left to right), and inside the positive and negative z-axis shell of the fullerene, $\rho_{f+}$ and $\rho_{f-}$ (second and third column).
}
\end{figure}

\begin{figure}
\begin{center}
\includegraphics[width=\textwidth]{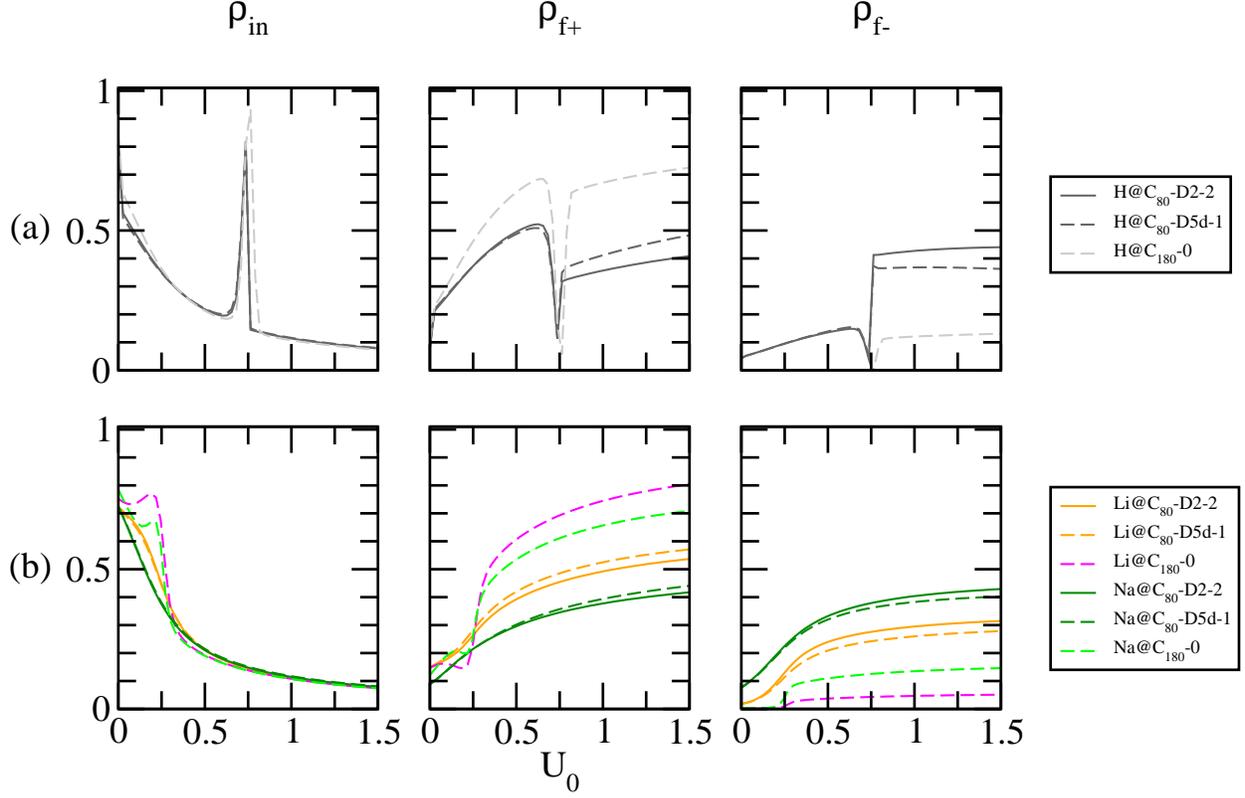}
\end{center}
\caption{  \label{frhon2} (color online) Probability of finding the valence electron in the excited state, inside the fullerene cage $\rho_{in}$ (first column from left to right), and inside the positive and negative z-axis shell of the fullerene, $\rho_{f+}$ and $\rho_{f-}$ (second and third column), for the hydrogen case (top) and the lithhium an sodium case (bottom row).
}
\end{figure}

With the aim of clarifying these features, we calculated as a function of $U_{0}$ the  probability of finding the valence electron inside the fullerene cage, 

\begin{equation}
\label{erhoin}
\rho_{in} \,=\,  \langle \Psi(\vec{x}) \vert \Psi(\vec{x}) \rangle  _{V_{f_{in}}} \,=\, \int_{0}^{\pi} \int _{0} ^{R(\theta)-\Delta} \, u^{2} (r, \theta) \, dr \, \sin  \theta \, d\theta  \,,
\end{equation}

\noindent where $V_{f_{in}}$ is the inner volume of the fullerene. We also calculated the probability of finding the valence electron in the positive upper ``hemisphere" of the fullerene shell, denoted by $\rho_{f+}$, 

\begin{equation}
\label{erhof}
\rho_{f+} \,=\,  \langle \Psi(\vec{x}) \vert \Psi(\vec{x}) \rangle  _{V_{f+}} \,=\, \int_{0}^{\frac{\pi}{2}} \int _{R(\theta)-\Delta} ^{R(\theta)+\Delta}  \, u^{2} (r, \theta) \, dr \, \sin  \theta \, d\theta \,,
\end{equation}

\noindent where $V_{f+}$ is the volume of the upper ``hemisphere" of the fullerene molecule. The probability of finding the valence electron in the lower ``hemisphere" of the fullerene $\rho_{f-}$ is defined in a similar manner. These probabilities are shown in Fig.  \ref{frhon1} for the ground state and in Fig. \ref{frhon2} for the first excited state.

It is clear that the valence electron of $Li$ and $Na$ becomes localized in the fullerene shell as soon as the interaction with the fullerene is non-zero, and this localization takes place in a soft and monotonic way. 

The studied hydrogen electronic states show a different behaviour. The ground state localization occurs in a step-like way at $U_{0} \sim 0.73$, as one can see from the staggered curves of $\rho_{in}$, $\rho_{f+}$ and $\rho_{f-}$ in Fig.  \ref{frhon1}. The mentioned re-confinement of the excited state in a narrow $U_{0}$ range is exposed in the high peak of $\rho_{in}$ (Fig. \ref{frhon2} (a)) for  $U_{0} \sim  [0.6,0.76]$ in the case of $C_{80}$ isomers and in the  $[0.65,0.82]$ range for $C_{180}$. As expected, for these values of $U_{0}$, $\rho_{f+}$, and $\rho_{f-}$ undergo an appreciable decrease.

As it is shown in Fig.  \ref{frhon1} and \ref{frhon2}, the main difference in electron localization is between $C_{80}$ isomers and $C_{180}$. For $C_{180}$, the nucleus of the enclosed atom locates at the largest distance from the lower ``hemisphere" of the fullerene; as a consequence, the valence electron ground state for all the elements becomes localized only in $f+$ (Fig. \ref{frhon1}), and the first excited states ($p$ wave for $U_{0}=0$) becomes localized mostly in $f+$ and shows poor localization in $f-$ (Fig. \ref{frhon2}). 

For $C_{80}$ cases, since the $Na$ nucleus locates the closest to the geometrical center of the fullerene, its states show a very equilibrated localization, which is more pronounced for $n=2$ (Fig. \ref{frhon2} (b)). The valence electron of $Li$ in the ground state shows a complete localization in $f+$, and as is expected given the symmetry of the initial state, the excited state displays a more equilibrate localization between $f+$ and $f-$.  


\begin{figure}
\begin{center}
\includegraphics[width=6.cm]{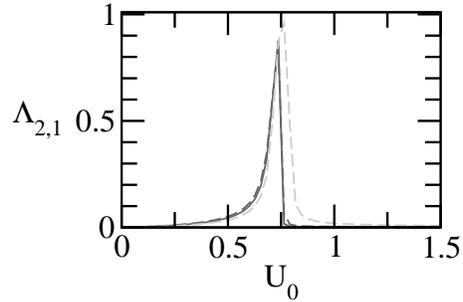}
\end{center}
\caption{  \label{fcin2_H} (color online) Projection of the first excited state onto the ground atomic free state for $H@C_{80}-D2-2$ (dark gray full line) $H@C_{80}-D2-2$ (dark gray dash line) and  $H@C_{180}-0$ (light gray dash line).
}
\end{figure}

Now, one could ask if the re-confinement of the excited state in the hydrogen case is a re-confinement into the atom or just into the inner volume of the fullerene. To answer this question we calculated the projections onto the free atomic states \textit{i.e.} without interaction with the fullerene ($U_{0}=0$), 

\begin{equation}
\label{eciproj}
\Lambda_{n,n_{0}} \,=\,  \langle \Psi_{n}(U_{0}) \vert \Psi_{n_{0}}(U_{0}=0) \rangle \,,
\end{equation}

\noindent where $n_{0}=1,2$ denote the ground and excited free atomic states. The projection of the first excited state of $H$ onto the ground atomic free state is displayed in Fig. \ref{fcin2_H}, which shows that the re-localization of the $H$ excited state is a re-confinement into the atom in a state that is quite similar to the atomic free ground state, and not only the return of the electron to the inner fullerene cage. 

Finally, we performed an analysis of the mixing of different angular partial waves. For this purpose we followed Kang \textit{et al.} \cite{kang2006} and calculated the partial wave-weights of the ground and excited states.

From the normalization condition of the wavefunction, using Eq. (\ref{eu_BS}), and expanding the angular B-Spline into Legendre Polynomials, 

\begin{equation}
\label{ewl3}
B_{j,\tilde{k}}(\theta)\,=\, \sum_{l} A_{l}^{j,\tilde{k}}\,P_{l}(\theta) \,,
\end{equation}

\noindent it is possible to define a partial wave overlap matrix, with elements given by, 

\begin{equation}
\label{ewl5}
\tilde{S}^{(l)} _{\left\lbrace i ,\, j \right\rbrace, \left\lbrace  \tilde{\imath}, \, \tilde{\jmath} \right\rbrace} \;=\; \left\lbrace  \int_{0}^{r_{max}} B_{i,k}(r) \, B_{\tilde{\imath},k}(r) \,dr  \right\rbrace \; \left\lbrace \frac{2}{2l+1} \, A_{l}^{j,\tilde{k}} A_{l}^{\tilde{\jmath},\tilde{k}}  \right\rbrace \,,
\end{equation}

\noindent and, based on this definition, the partial wave-weights of the state are obtained,

\begin{equation}
\label{ewl6}
W_{l} \;=\; \vec{c} \,\, \tilde{S}_{l} \,\, \vec{c} \,.
\end{equation}

\begin{figure}
\begin{center}
\includegraphics[width=14.cm]{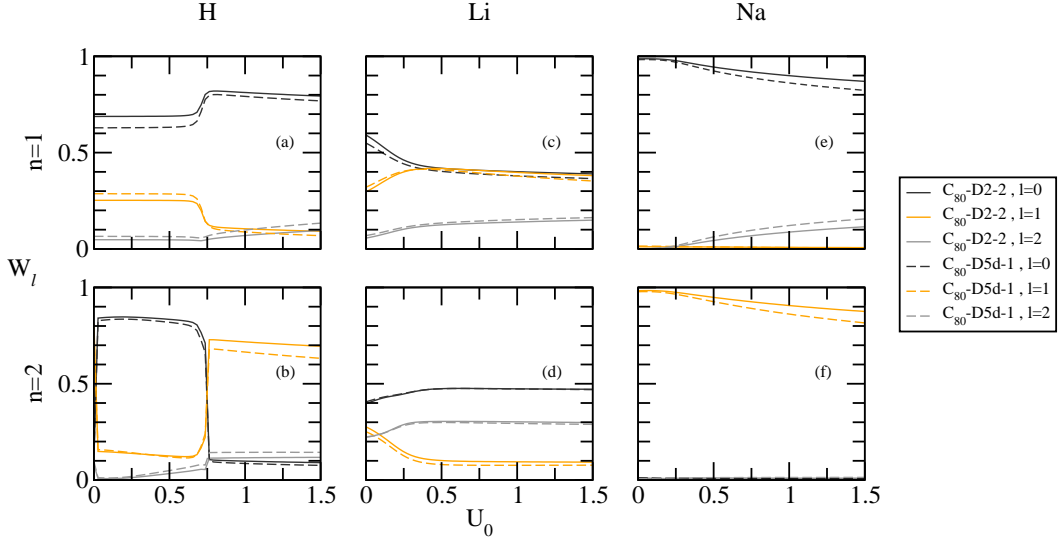}
\end{center}
\caption{  \label{fpl_C80} (color online) Partial weights of $s$, $p$, and $d$-waves for $C_{80}$ isomers. The different elements are in columns ($H$, $Li$ and $Na$ from left to right) and the ground and excited states are displayed in the rows from top to bottom.
}
\end{figure}

The partial weights of $s$, $p$ and $d$-waves as a function of $U_{0}$ are shown in Fig.  \ref{fpl_C80} for the $C_{80}$ isomers.

As expected, the larger the distance between the nucleus of the enclosed atom and the geometrical center of the fullerene is, the more the partial waves get mixed. Therefore, the $Li$ states present a higher mixing of partial angular waves (see Fig. \ref{fpl_C80} (c) and (d)) while $Na$ shows the purest wave-states; the valence electron ground state of $Na$ is mostly an $s$-wave (see Fig. \ref{fpl_C80} (e)) and its excited states, a $p$-wave (Fig. \ref{fpl_C80} (f)). 

All of the hydrogen weights exhibit a quasi-critical behavior at $U_{0} =U_0^{(c)}\sim 0.73$ (see Fig.  \ref{fpl_C80} (a) and (b)). In particular, the first excited state of $H@C_{80}$ switches its principal partial angular weight from $s$ to $p$ as can be seen in Fig. \ref{fpl_C80} (b). This means that by increasing from zero the values of $U_{0}$, the electron becomes localized in the fullerene as an $s$-wave. After that, for $U_{0}$ values in a small interval around the anticrossing value $U_{0}^{(c)}$, the electron is confined into the atom and relocalized in the fullerene, but this time as a $p$-wave.
 

\section{Summary and conclusions}

In this work we described how the valence electron of $H$, $Li$ and $Na$ atoms enclosed by three different fullerene molecules ($C_{80}-D2-2$, $C_{80}-D5d-1$ and $C_{180}-0$) localizes in the fullerene when the strength of the attractive molecule cage increases. 

We used the structure of the fullerene molecules to calculate the equilibrium position of the endohedrally atom as the minimum of the classical $(N+1)$-body Lennard-Jones potential. From LJ parameters (Table \ref{tljp}) we obtained that $Li$ atoms locate farther than $H$ and $Na$ atoms in all cases. 

Once the position of the enclosed atom was determined, the fullerene molecules were modeled by a short range attractive shell according to molecule symmetry, and the enclosed atoms were modeled by an effective one-electron potential. $C_{180}$ was modeled by a spherical shell, while both $C_{80}$ isomers were described by an ellipsoidal shell with proper parameters. 

We found that the localization of the valence electron on the fullerene shell depends on the distance between the nucleus of the enclosed atom and the geometrical center of the fullerene, as well as on the symmetry of the atomic free states \textit{i.e.} without interaction with the fullerene cage. As we expected, the main differences between fullerene molecules were observed when comparing $C_{180}$ and $C_{80}$ isomers.

The $Li$ and $Na$ valence electron becomes localized in the fullerene shell continuously and monotonically as soon as the interaction with the fullerene is non-zero. Thus, $Li$ and $Na$ atoms inside a fullerene cage present zwitterion-like states \textit{i.e.} the endohedral compound has the form $Li^{+}@C_{N}^{-}$ and $Na^{+}@C_{N}^{-}$.  

The hydrogen electronic states present a different behavior. In order to understand it, we analyzed the electron probability density, the projections onto free atomic states and the weights of partial angular waves. The $H$ ground state remains unaltered until $U_{0}$ values get over $\sim 0.73$; beyond this value the electron becomes localized in the fullerene shell in a step-like way. Therefore, the endohedral compound in the case of hydrogen is formed by a neutral atom inside a neutral fullerene $H@C_{N}$. 

The first excited $H$ state presents a quite interesting phenomenon which coincides with $U_{0}$ values for the anticrossing energy levels. Increasing $U_{0}$ while still remaining under the anticrossing values, the electron becomes localized in the fullerene as an $s$-wave. For anticrossing values the electron is confined into the atom in a state similar to the atomic free ground state. When $U_{0}$ takes values slightly above the anticrossing, the electron is re-localized in the fullerene shell as a $p$-wave. This switching of the principal partial angular weight could be a possible way to define the avoided crossing $U_{0}$ value.


We would like to make some final remarks on one of the possible applications of the  simple approach used in the present work. So far, Light Induced Conical Intersection (LICI) as first proposed by M. Sindelka, N. Moiseyev and L.S. Cederbaum was applied to diatomic molecules in running and standing laser waves \cite{sindelka2011}. Within the present approach it is possible to study the coupling between nuclear and electronic modes in endohedral compounds when a conical intersection \cite{cederbaum2004} between electronic levels is induced by laser waves. Work is in progress in this direction.

\section*{Acknowledgements}

We acknowledge  SECYT-UNC and CONICET for partial financial support of this project. E.C. would like to thank Alvaro Cuestas for an exhaustive reading of the manuscript.

\end{document}